\documentclass[letterpaper]{article} 
\usepackage{aaai25}  
\usepackage{times}  
\usepackage{helvet}  
\usepackage{courier}  
\usepackage[hyphens]{url}  
\usepackage{graphicx} 
\usepackage{natbib}  
\usepackage{caption} 
\usepackage{algorithm}
\usepackage{algorithmic}

\usepackage{newfloat}
\usepackage{listings}

\usepackage{booktabs}
\usepackage{array}
\usepackage{amsmath}
\usepackage{amssymb}
\usepackage{xcolor}
\usepackage[most]{tcolorbox}

\definecolor{mybrown}{RGB}{152,69,14}
\definecolor{lightbrown}{RGB}{254,249,245}

\definecolor{myblue}{RGB}{46,117,181}
\definecolor{lightblue}{RGB}{249,251,253}

\definecolor{mypurple}{RGB}{120,129,255}
\definecolor{lightpurple}{RGB}{245,242,255}

\definecolor{mylakeblue}{RGB}{65,170,255}
\definecolor{lightlakeblue}{RGB}{245,250,255}

\definecolor{mygptgreen}{RGB}{115,170,155}
\definecolor{lightgptgreen}{RGB}{240,245,245}

\definecolor{mycat}{RGB}{250,195,70}
\definecolor{lightcat}{RGB}{250,240,225}

\definecolor{myprompt}{RGB}{240,110,70}
\definecolor{lightprompt}{RGB}{250,240,235}

\newcommand{\appendixheading}{
  \vspace*{0.5em}
  \begin{center}
    {\Large\bfseries Appendix}
  \end{center}
  \vspace{0.3em}
}

\newcommand{\appendixsection}[1]{
  \refstepcounter{section}
  \vspace{1.0em}
  \noindent
  {\normalsize\bfseries Appendix~\thesection.\quad #1}
  \par
  \vspace{0.25em}
  \noindent\rule{\linewidth}{0.35pt}
}

\DeclareCaptionStyle{ruled}{labelfont=normalfont,labelsep=colon,strut=off} 
\floatstyle{ruled}
\newfloat{listing}{tb}{lst}{}
\floatname{listing}{Listing}
\title{Agent4Edu: Generating Learner Response Data by Generative Agents\\for Intelligent Education Systems}
\author {
	Weibo Gao\textsuperscript{\rm 1},\;
	Qi Liu\textsuperscript{\rm 1,2}\thanks{ Corresponding Author.},\;
	Linan Yue\textsuperscript{\rm 1},\;
	Fangzhou Yao\textsuperscript{\rm 1},\;
	Rui Lv\textsuperscript{\rm 1},\;
	Zheng Zhang\textsuperscript{\rm 1},\\
	Hao Wang\textsuperscript{\rm 1},\;
	Zhenya Huang\textsuperscript{\rm 1,2}
}
\affiliations {
	\textsuperscript{\rm 1} State Key Laboratory of Cognitive Intelligence, University of Science and Technology of China\\
    \textsuperscript{\rm 2} Institute of Artificial Intelligence, Hefei Comprehensive National Science Center \\
	\{weibogao, lnyue, fangzhouyao, lvrui2018, zhangzheng\}@mail.ustc.edu.cn \\ \{qiliuql, wanghao3, huangzhy\}@ustc.edu.cn
}

\begin{document}

\maketitle

\begin{abstract}
Personalized learning represents a promising educational strategy within intelligent educational systems, aiming to enhance learners' practice efficiency. However, the scarcity of offline practice response data (e.g., answer correctness) and potential biases in human online practice create a significant gap between offline metrics and the actual online performance of personalized learning services. To address this challenge, we introduce Agent4Edu, a novel personalized learning simulator leveraging recent advancements in human intelligence through large language models (LLMs). Agent4Edu features LLM-powered generative agents equipped with learner profile, memory, and action modules tailored to personalized learning algorithms. The learner profiles are initialized using real-world response data, capturing practice styles and cognitive factors. Inspired by psychology theory, the memory module records practice facts and high-level summaries, integrating reflection mechanisms. The action module supports various behaviors, including exercise understanding, analysis, and response generation. Each agent can interact with personalized learning algorithms, such as computerized adaptive testing, enabling a multifaceted evaluation and enhancement of customized services. Through a comprehensive assessment, we explore the strengths and weaknesses of Agent4Edu, emphasizing the consistency and discrepancies in responses between agents and human learners.
\end{abstract}

%

\section{Introduction}
Intelligent education platforms like \textit{Coursera.com} and \textit{LeetCode.com} provide a rich array of learning resources, such as courses and exercises, within a flexible online environment. The accessibility and convenience of these platforms have attracted a growing number of learners. A key online learning activity is ``practice'', where learners independently select and answer exercises. The platforms record their responses, such as the correctness of their answers. By analyzing response data, many personalized learning services, such as exercise recommendations, knowledge tracing, and computerized adaptive testing, can be tailored to meet each learner's specific needs, enhancing the learning process and increasing learner satisfaction. For instance, on \textit{LeetCode}, analyzing a learner's historical programming experiences allows the platform to recommend exercises of appropriate difficulty levels, thus optimizing learning gains.

The effectiveness of personalized learning services hinges on the availability of high-quality response data for the corresponding algorithm training. However, the scarcity of offline response data and potential biases in its correlation with online practice introduces a significant gap between offline metrics and actual online performance. This discrepancy impedes the integration of research with real-world applications.
To bridge this gap, a promising approach is to simulate learner response data. Imagine an online platform equipped with a configurable simulation system that faithfully captures human learners' response patterns while seamlessly interacting with personalized learning algorithms. Such a simulator undoubtedly has the potential to revolutionize the traditional research paradigm in intelligent education, providing innovative avenues for response data collection, personalized algorithm development and evaluation. 

Several approaches to simulating learner response data have been proposed and have achieved notable success~\cite{piech2015deep,zhao2023simulating}. However, two major limitations exist in current approaches:
(1) \textbf{Simplified Simulations.} Most existing studies predict learners’ responses (e.g., correct or incorrect answers) without considering the detailed answer processes by which humans use their knowledge to understand, analyze, and solve problems. Hence, these simulations may lack reliability and interpretability.
(2) \textbf{Dependency on Real Response Data.} An ideal simulator should be capable of simulating learner responses even when real-world datasets are insufficiently available, thereby enhancing its applicability. However, current methods require high-quality real-world data to train the simulation strategy. As a result, these methods can only generate learner response data similar to existing real-world datasets and struggle to generalize to more challenging scenarios, such as zero-shot simulations.

Recent advancements in large language models (LLMs) have demonstrated remarkable capabilities in autonomous interaction and decision-making~\cite{brown2020language,ouyang2022training,yue2023fedjudge, jin2023data, long2024generative}. These advancements underscore the potential of leveraging LLM-powered agents to simulate human social behaviors, such as daily life in Smallville~\cite{park2023generative} and software development~\cite{qian2023communicative}. LLM-based user simulators possess rich pre-trained knowledge and human-like intelligence, enabling them to perceive and simulate intricate human practice processes. Furthermore, their in-context learning ability allows LLMs to perform zero-shot simulations with minimal reliance on real-world data~\cite{wang2023recagent}. Consequently, LLM-based generative agents present a promising approach for addressing the current limitations of learner response simulators.

\begin{figure*}[t]
	\centering
	\scalebox{0.25}
	{\includegraphics{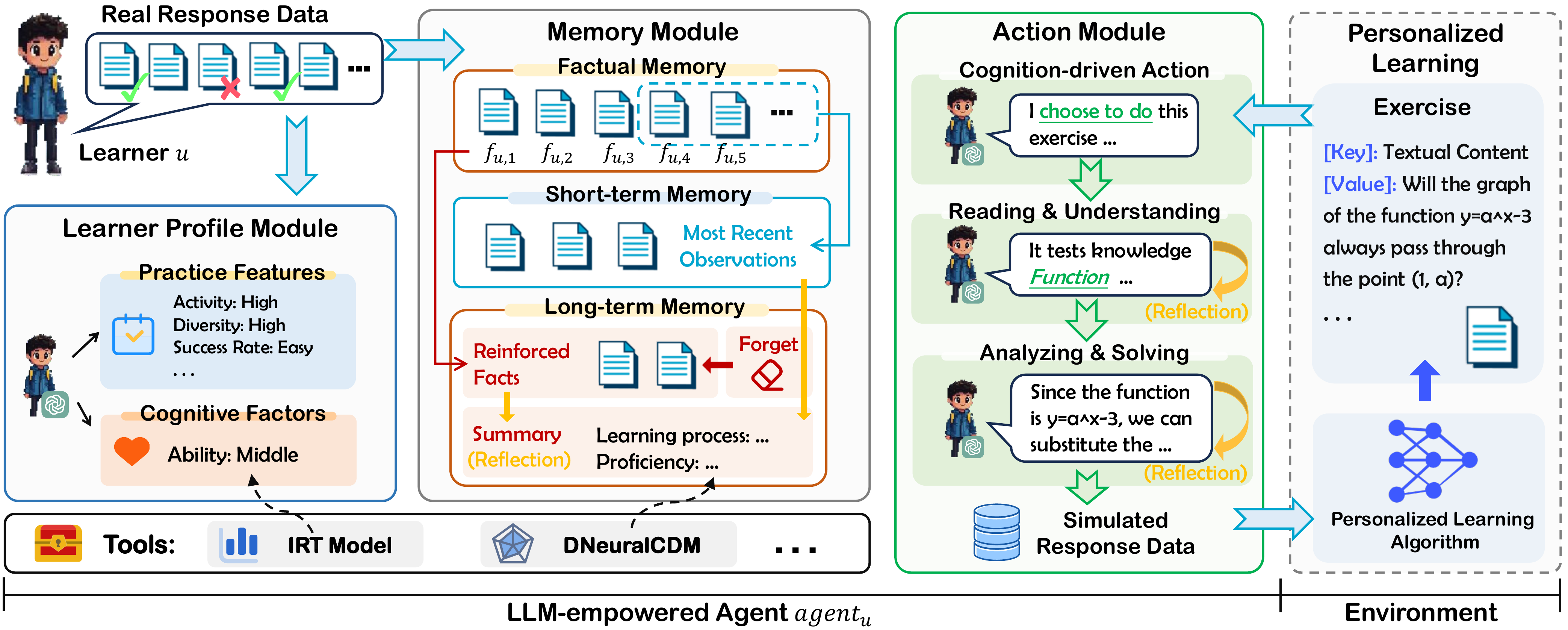}}
	\caption{The overall framework of Agent4Edu.}
	\label{fig:model}
\end{figure*}

In this paper, we introduce \textbf{Agent4Edu}, a personalized learning simulator designed for intelligent educational systems, comprising two key components: an \textbf{LLM-powered generative agent} and a \textbf{personalized learning environment} (see the framework in Figure~\ref{fig:model}).
From a learner perspective, the LLM-powered generative agent is responsible for simulating learners' response data by capturing their response patterns and inferring problem-solving actions. Each agent is initialized based on available learner response data and consists of three modules: a learner profile, memory, and action module. The learner profile module stores learners' past practice styles (e.g., \textit{activity}) and cognitive factors (e.g., \textit{ability}), aligning with human learners' learning status. The memory module, inspired by psychological theories~\cite{baker2001basics} and human learning mechanism~\cite{wang2023user}, records past practice experiences and summarizes learning status through reflections. This facilitates coherent observations, monitors \textit{knowledge proficiency evolution}, reinforces memory, and simulates human forgetting.
The action module enables agents to choose, understand, analyze, and solve exercises recommended by personalized learning algorithms, leading to more reliable and interpretable response generation.
Our agent can also utilize tools, such as employing the psychological IRT model~\cite{baker2001basics} to assess \textit{ability} within the Profile module and using DNeuralCDM~\cite{wang2023dynamic} to trace \textit{knowledge proficiency evolution} within the Memory module.
From a personalized learning perspective, the learning environment can be configured with any personalized learning algorithm, allowing agents to interact directly and simulate a real learning environment.
Notably, despite extensive research on simulating user behavior with generative agents, we are the first to focus specifically on educational scenarios to generate response data for individual learners.

Our main contributions are summarized as follows:
\begin{itemize}
\item We develop Agent4Edu\footnote{The code, data, and appendix are publicly available at \url{https://github.com/bigdata-ustc/Agent4Edu}.}, a personalized learning simulator that leverages LLM-powered generative agents to simulate human learners' response data as well as demonstrate the practice process. Additionally, the agent interacts with personalized learning environments to evaluate and improve intelligent tutoring algorithms.
\item Our generative agents, featuring profile, memory, and action modules specifically designed for ``Education'', can not only generate response data but also accurately simulate human choices, understanding, analysis, and problem-solving for exercises, outperforming existing learner simulation methods.
\item To systematically evaluate Agent4Edu, we conduct comprehensive experiments from both the agent and personalized learning perspectives. From the agent perspective, we assess the consistency between the agents and human learners. From the learning perspective, we evaluate and improve personalized learning algorithms for computerized adaptive testing, based on generative agents and simulated data. Extensive experimental results demonstrate the effectiveness of Agent4Edu.
\end{itemize}

\section{Related Work}
\textbf{Learner Response Data Simulation} 
Learner Simulation aims to address the shortage of high-quality practice data in intelligent educational systems and has been applied in numerous previous studies~\cite{zhao2023simulating,yao2024adard}. Memory-based~\cite{reddy2017accelerating} relies on manually crafted rules to predict learners' responses or memory behavior. EERNN~\cite{su2018exercise} and KES~\cite{liu2019exploiting} utilize RNN-based models to forecast learners' performance. DAISim~\cite{zhao2023simulating} constructs learner simulations as Markov decision processes, simultaneously considering learners' long and short-term question-answering patterns. However, the memory-based simulator is overly simplistic and cannot simulate complex interactions. Most other learner simulators simplify the student answering process and face challenges in conducting zero-shot simulations due to their reliance on data. In this paper, we employ an LLM-powered agent to simulate the student practice process, addressing these limitations.

\noindent\textbf{Personalized Learning Services}
Intelligent educational systems offer learners personalized learning services, including Computerized Adaptive Testing (CAT)~\cite{chang1996global}, exercise recommendation~\cite{huang2019exploring} and learning path suggestions~\cite{liu2019exploiting}, to help learners enhance their skills. 
In this work, we select the representative and popular CAT services as our personalized learning scenarios for our study and experiments. 
CAT is an advanced educational measurement method that evaluates the knowledge level of examinees in minor exercises, which has been widely used in various standardized tests (e.g., GMAT and GRE)~\cite{zhuang2024bounded, bi2020quality,lord2012applications,chang1996global}. However, current CAT models require high-quality practice data to train a cognitive diagnosis model for evaluating learner ability or knowledge proficiency, which is often challenging to gather. Therefore, in this paper, we employ the CAT service within learning systems to assess the quality of the data generated by our agents. Additionally, we investigate the potential for enhancing CAT models using simulated data.

\noindent\textbf{LLM-based Agents}
LLM-based generative agents demonstrate the remarkable capabilities to perceive their environment, make decisions, and take actions, thus, emerging a substantial amount of research~\cite{wang2024survey}. The development of generative agents~\cite{park2023generative}, designed with profile, memory, action, and reflective capabilities, represents pioneering work in simulating human daily life. Within this general framework, agents tailored to specific tasks~\cite{qian2023communicative,wu2023autogen,wang2023voyager,huang2023recommender,zhang2023agentcf,zhang2023generative} and simulations~\cite{gao2023s,wang2023user,park2023generative,liu2023training,wang2023recagent} have been constructed. Recent research highlights bringing generative agents to educational settings~\cite{li2024bringing,dan2023educhat,kieser2023educational}. For example, \cite{qadir2023engineering,rahman2023chatgpt} conclude the applications of ChatGPT to engineering education. \cite{baidoo2023education} focus on the literature review over the published paper. SocraticLM~\cite{liusocraticlm} embodies a ``Thought-Provoking'' teaching paradigm, engaging students in active problem-solving, akin to a real classroom teacher. The most relevant part of our work is EduAgent~\cite{xu2024eduagent} which utilizes LLM-based agents to simulate learners studying PowerPoint presentations and videos, predicting their quiz outcomes to assess performance. However, this approach relies on expert-annotated cognitive factors to initialize agents, disregarding the understanding and analysis of exercises. In contrast, our Agent4Edu extracts cognitive factors from data using tools and captures practice styles, allowing it to simulate the detailed exercise understanding and analysis process and interact effectively with personalized learning algorithms.

\section{Agent4Edu}
Agent4Edu is a personalized learning simulator, aimed at accurately simulating learners' response data and facilitating responsive personalized learning algorithms. It contains two key components: (1) LLM-powered generative agents that capture learners' practice patterns and cognitive preferences to simulate their response, and (2) a personalized learning environment that interacts with agents to support accurate and interpretable evaluations and improvements of mainstream intelligent algorithms (e.g., computerized adaptive testing).
The framework of Agent4Edu is illustrated in Figure~\ref{fig:model}. All the prompts are listed in Appendix~C.

\subsection{Task Formulation}
Suppose there are $|U|$ learners, $|
E|$ exercises in an intelligent educational system.
For a learner $u \in U$, his/her response data are denoted as a time-ordered set $l_{u}=\{(e_1,c_{e_1},y_{u,e_1}),(e_2,c_{e_2},y_{u,e_2}),\dots,(e_n,c_{e_n},y_{u,e_n})\}$, where $e_i \in E$ represents the exercise that learner $u$ practiced at step $i$, and $y_{u,e_i}$ is $u$'s response to exercise $e_{i}$, which is usually denoted as a binary value, i.e., if learner $u$ answers $e_i$ correctly, $y_i=1$ otherwise $y_i=0$. $c_{e}$ denotes textual information of each exercise $e \in E$, e.g., textual content and corresponding knowledge concepts. We provide $c_{e}$ in a $<key,value>$ form, as the example in Figure~\ref{fig:model}.

Based on the above conditions, the simulator’s overarching goal is to accurately generate the human learners' future response data on unseen exercises by faithfully distilling their learning patterns and cognitive preferences from historical response data.
Please note that existing personalized learning algorithms usually assume that learners only submit each exercise once, so repeated submission is not considered in our simulation.

\subsection{LLM-powered Agent}
The generative agent in Agent4Edu uses LLM as its foundational architecture, enhancing its functionality tailored for the personalized learning scenario through three specialized modules: learner profile, memory, and action modules. To mimic actual personalized practice responses akin to humans, we construct an individual agent $agent_{u}$ for each learner $u$. Each agent integrates a learner profile module aimed at reflecting personalized practice patterns and cognitive factors. Additionally, each agent is equipped with a memory module designed to store past practice records and summarize high-level ideas. To simulate learner practice behavior more cohesively, the agent is also equipped with an action module.

\subsubsection{Learner Profile Module} \label{sec:profile}
The learner profile module represents some overall learning features of human learners, which are typically stable and derived from long-term learning experiences. We configure each agent $agent_{u}$'s profile based on its corresponding learner $u$'s response data\footnote{Note that if zero-shot simulations are performed and user data is unavailable, the profile needs to be randomly generated.}. Each agent's initial configuration is divided into two categories: explicit practice styles and implicit cognitive factors.

Practice styles are statistical features explicitly derived from the available practice record \( l_u \) of each learner \( u \), such as learning activity~\cite{baker2001basics, gao2021rcd}, practice diversity~\cite{bi2020quality}, success rate, and preference. Activity indicates learners' enthusiasm for learning and provides clues for simulating their practice behaviors. For example, learners with higher enthusiasm for learning usually perform better. Mathematically, the activity level of learner \( u \) is defined as \( P_{\text{act}}^u = \frac{|l_u|}{|E|} \). Practice diversity reflects the knowledge coverage practiced by learners, represented as \( P_{\text{div}}^u = \frac{|K_u|}{|K|} \), where \( |K_u| \) is the number of knowledge concepts practiced by learner \( u \). Higher diversity indicates greater curiosity in learners. Success rate correlates with the probability of learners answering questions correctly, making it another essential characteristic. The success rate for learner \( u \) is mathematically represented as \( P_{\text{suc}}^u = \frac{\sum_{y_{u,e_i} \in l_u} y_{u,e_i}}{|l_u|} \). Preference refers to the knowledge concepts that learners practice most frequently.

Cognitive factors are implicit features studied in psychology~\cite{baker2001basics, chen2024disentangling}, which significantly impact learner \( u \)'s practice performance. We select problem-solving ability and knowledge proficiency~\cite{cheng2024dygkt} for this study. Problem-solving ability is assumed to be stable during the learning process, while knowledge proficiency typically evolves with learning progress~\cite{huang2020learning}. Therefore, in the profile module, we only configure the ability factor \( P_{\text{ab}}^{u} \), with knowledge mastery being considered in the subsequent memory module. To obtain implicit ability, we assign a psychological IRT model~\cite{baker2001basics} trained on the observed learner response records, as the tool for the agent, allowing it to infer each learner \( u \)'s ability factor from the response data \( l_u \). The training and use of the IRT tool are detailed in Appendix~B.

Notably, we segment the values of each of the above features into several tiers in order to better prompt the generative agent inspired by~\cite{wang2023user}. 
For a detailed exposition,
refer to Appendix~A.1.
Additionally, to ensure broad applicability and protect privacy, certain personal identifiers (such as name, gender, age, and occupation) are intentionally anonymized in this work~\cite{zhang2023generative, li2023understanding}. While these attributes may help shape other types of agents, they are not primary factors affecting practice performance in education. Our approach, based on both behavioral practice styles and psychological cognitive settings, can support a comprehensive representation of real learners.

\subsubsection{Memory Module} \label{sec:memory}
The Memory module allows the LLM-based agent \( agent_{u} \) to observe and summarize its corresponding learner \( u \)'s past practice experiences step by step. This module provides insightful clues to the agent for response simulation on unseen exercises. We follow the human learning mechanism~\cite{atkinson1968human, cowan2008differences, huang2020learning, wang2023user} to design three types of memories for each agent: \textit{factual memory}, \textit{short-term memory}, and \textit{long-term memory}. Each memory is initially set to empty.

\noindent$\small{\bullet}$~\textit{Factual Memory}:
In our simulation, factual memory is defined as the true learner's past response records (i.e., \textit{observations}). When the agent obtains a new response record of learner $u$ at step $i$, i.e., $l_{u,i}=(e_i,c_{e_i},y_{u,e_i})$, the response record is transmitted to the factual memory for processing.

Inspired by human learning mechanisms, if an agent repeatedly practices similar questions or knowledge, their memory is strengthened~\cite{huang2020learning}. Therefore, we introduce an additional counter \( f_{u,i} \) (initially set to 1) for each record \( l_{u,i} \) in factual memory to track the number of times it has been reinforced, a simple yet effective method that has been successfully used in user preference simulation~\cite{wang2023user}. Formally, for each agent $agent_u$, assume it has observed $n$ factual memories is \(M_u = \{l_{u,1}, l_{u,2}, \ldots, l_{u,n}\}\), then it is allowed to receive a new response record \(l_{u,n+1}\). We first calculate the similarity between \(l_{u,n+1}\) and each existing factual memory \(l_{u,i}\) in the current memory $M_u$.
The similarity between records can be defined as a metric that can be evaluated by LLMs, cosine similarity between text vectors, and other similar measures. In this case, we use the similarity relationships between the knowledge concepts involved in the records for the calculation. Specifically, we employ the statistical tool\footnote{\url{https://github.com/bigdata-ustc/RCD}} released by RCD~\cite{gao2021rcd} to determine whether two knowledge concepts are similar. If there is a similarity between the knowledge concepts involved in two records, the two records are considered similar.
For similar records, we increment the counter for \(l_{u,i}\) by 1 (i.e., $f_{u,i} \leftarrow f_{u,i}+1$), indicating that it has been reinforced by \(l_{u,n+1}\), and then add \(l_{u,n+1}\) to factual memory; otherwise, \(l_{u,n+1}\) is directly recorded without any reinforcement.
After processing and saving a new response record, factual memory triggers updating \textit{short-term} and \textit{long-term memories}.

We emphasize that the agent can only save response records into factual memory but cannot directly retrieve it, thereby allowing the retention of all exercise textual information and responses without being constrained by the LLM's context length limitations.

\noindent$\bullet$~\textit{Short-term Memory}:
Human short-term memory refers to the recent and temporary information that can be retained and recalled over a relatively brief period~\cite{atkinson1968proposed}. Therefore, in our simulation, short-term memory is employed to retain the details of the agent's most recent observed $s$ records. Assuming the current factual memory of agent $agent_u$ is \(M_{u} = \{l_{u,1}, l_{u,2}, \ldots, l_{u,n}\}\), then the short-term memory storage is defined as \(M_{u,short} = \{l_{u,n-s+1}, \ldots, l_{u,n}\}\).

\noindent$\bullet$~\textit{Long-term Memory}:
Long-term memory is formed through the reinforcement of memories from repeated practice and self-reflection inspired by to human long-term memories~\cite{matelsky2023large}. It possesses a wide receptive field, allowing it to retain information observed long ago and generate high-level insights. We design the long-term memory using three types of information:
(1) \textbf{Reinforced Facts}: During each update of long-term memory, the agent $agent_u$ first goes through the current factual memory $M_{u}$. When the count $f_{u,i}$ of a record $l_{u,i}$ exceeds a preset threshold \(F\), indicating that the memory has been reinforced \(F\) times, it is converted into long-term memory.
(2) \textbf{Learning Process Summary}: We utilize the LLM embedded in the agent to summarize the agent's learning status from both short-term and long-term memories by \textit{Memory Reflection}. Each step of the summary replaces the previous summary.
The summary consists of linguistic descriptions of the practice process and new insights from the agent itself. It overlooks practice details to filter out noise, irrelevant content, or potentially misleading information. Furthermore, compressing memory conserves significant space and enhances operational efficiency. (3) \textbf{Knowledge Proficiency}: We allow the agent to use an optimized DNeuralCDM~\cite{wang2023dynamic} based on the observed learner response data as a tool to obtain the learner's dynamic proficiency (segmented into several tiers) evolution of specific knowledge concepts after each step of practice. The knowledge proficiency is a kind of dynamic cognitive factor significantly reflecting human responses in education~\cite{piech2015deep,wang2024survey_cd}. The training and use of DNeuralCDM are given in Appendix~B.

Additionally, each factual record in long-term memory may be \textbf{forgotten} following the human forgetting curve theory~\cite{averell2011form,huang2020learning} that human memory decay starts rapidly and then gradually slows over time. We define a forgetting function associated timestamp $i$ and current observed step $n$, i.e., $g(l_{u,i})=\frac{1}{1+e^{-(n-i)}}$, to simulate human learners' forgetting. For each factual record in the long-term memory $M_u$, it is forgotten if $g(l_{u,i})$ exceeds a predetermined threshold $\lambda$ and its reinforcement frequency $f_{u,i}$ in factual memory is then reset as $1$.

Overall, the factual response records are specific, while learning memory summaries are more general. By combining them, the agent can accurately perceive the learner's practice process.
Please note that traditional simulators~\cite{piech2015deep,zhao2023simulating} can be regarded as owning the short-term memory but no long-term memory.

To help agents interact with the personalized learning environment, we introduce three memory operations:

\noindent$\bullet$~\textit{Memory Retrieval}: This operation assists the agent in extracting related information from memory. We allow the agent to retrieve the short-term and long-term memories finding reinforced facts and conducting summary.

\noindent$\bullet$~\textit{Memory Writing}: The raw observations are firstly input into the factual memory as facts. Then the recent facts are stored in short-term memory and the reinforced facts are written into long-term memory.
    
\noindent$\bullet$~\textit{Memory Reflection}: This operation occurs exclusively within long-term memory containing two aspects of reflections: (1) \textbf{Summary Reflection} is performed to summarize high-level ideas based on short-term and long-term memories, and (2) \textbf{Corrective Reflection} is performed when the agent's action is inconsistent with the real learner, which will be introduced in Action Module.

\subsubsection{Action Module}
To equip the agent with learner profiles and memory modules, enabling it to exhibit human-like problem-solving behaviors and responses based on current observations, we design a specialized action module for each agent within Agent4Edu tailored for personalized learning. This module encompasses three main categories of actions:

\noindent$\bullet$~\textit{Cognition-driven Actions}:
In our simulation, personalized learning algorithms recommend one exercise to the agent at each step. The agent read the exercise's content and decides whether or not to practice it, based on current cognitive factors. If the exercise is too challenging relative to the agent's assessed ability and knowledge proficiency, the agent can opt to reject the recommended exercise.

\noindent$\bullet$~\textit{Reading and Understanding Exercises.}
Simulating the process of reading and understanding exercises, similar to how humans approach them, provides valuable and interpretable insights into the agents. During each practice session, the agent is first required to identify and describe a knowledge concept assessed by the current exercise. If the agent correctly matches the exercise's knowledge concept, it demonstrates an understanding of the exercise context akin to human learners. If the agent fails to do so, a corrective reflection is triggered to guide the agent towards the correct knowledge concept. This method reduces the risk of inaccuracies and ensures the agent's credibility in simulating learner response~\cite{zhang2023generative}.

\noindent$\bullet$~\textit{Analyzing and Solving Exercises.}
Analyzing and solving exercises are crucial aspects of the learning process. Unlike previous simulation methods that directly predict the learner's response in terms of answer correctness, our simulation requires the agent to emulate the learner's answering process, which enhances both interpretability and credibility. To simulate this complex answer process more effectively, we improve agent's reasoning ability through a chain-of-thought approach~\cite{wei2022chain}. Initially, the agent combines its profile and memories to formulate an initial solution idea for the exercise. Then, it writes the final answer to the exercise based on the solution idea. Afterwards, the agent predicts whether its answer is correct (i.e., performance prediction). If the predicted response does not match the real learner's response, a corrective reflection is triggered. Note that, if standard answers of exercises are available, a scoring program can be designed to directly assess the correctness of the agent's answer.

\subsection{Personalized Learning Scenarios}
Agent4Edu simulates agent and learning environment interaction (see Appendix~D for a case study). The learning environment is designed as a standalone module that incorporates a series of personalized algorithms. These algorithms can recommend exercises to agents based on their past practice data. For instance, our experiments utilize computerized adaptive testing (CAT) strategies~\cite{bi2020quality} for personalized learning. The module features an open interface, allowing researchers and practitioners to integrate external personalized learning algorithms seamlessly. This adaptability ensures that Agent4Edu serves as a versatile platform for comprehensive evaluations and the future collection of valuable learner response data.

\section{Experiment}
\paragraph{Dataset}
Our dataset, called \textit{EduData}, is provided by iFLYTEK Co., Ltd.
It comprises 18,045 time-ordered response records from 500 Chinese high school students in the subjects of mathematics and physics. Each record includes the exercise ID, correctness, and timestamp. There are 1,032 exercises and 458 knowledge concepts in total, with each exercise testing one knowledge concept. Additionally, to facilitate reasoning and reflection for LLM-based agents, the platform provider has furnished us with the textual content of the exercises. In the experiment, we translate all Chinese text of exercises into English.

\paragraph{Experimental Setup} \label{sec:setup}
We use GPT-3.5-turbo and GPT-4 through OpenAI’s API service~\footnote{The detailed GPT versions: GPT-3.5-turbo-1106 (up to Sep. 2021) and GPT-4-turbo (up to Dec. 2023).} to construct the agent for experimentation. When operating under the GPT-3.5-turbo configuration, all response data is utilized for experiments. Due to cost considerations, we simulate the task records of only 100 learners under the GPT-4 setting. The temperature parameter of GPT is 0 to avoid randomness.
Empirically, we set the short-term memory size to 5, the threshold $F$ for memory enhancement is 5, and the threshold $\lambda$ for forgetting in long-term memory is 0.99.
Note that, in our experiments, unless explicitly specified, the LLM used is GPT-3.5-turbo.
\begin{figure*}[t]
	\centering
	\scalebox{0.4}
	{\includegraphics{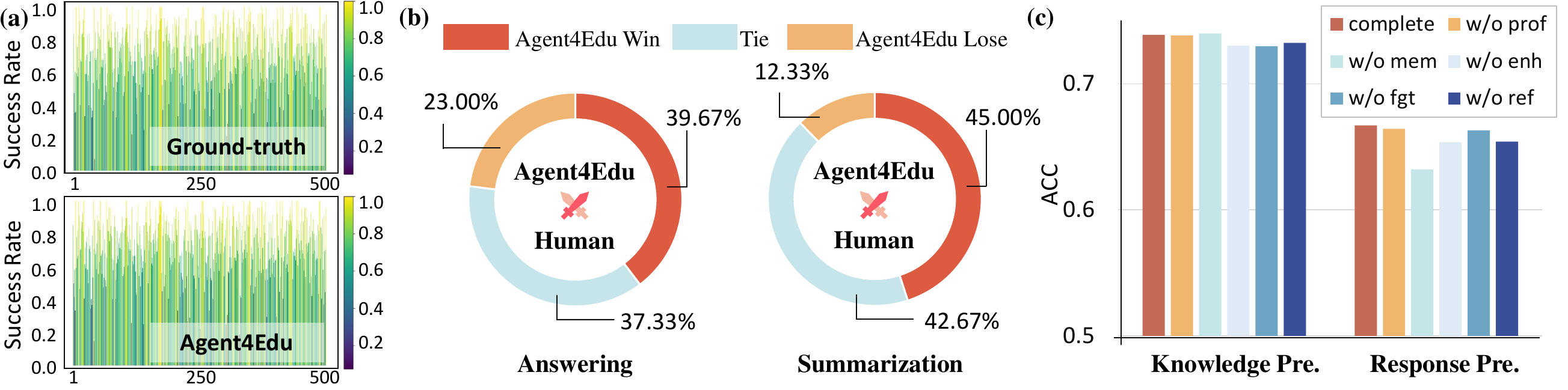}}
	\caption{(a) Comparison between the success rate distributions of ground-truth and agent-simulated response data. (b) Using LLMs as judges to identify whether the records of the agent’s simulations originate from real humans. (c) The ablation studies.}
	\label{fig:circle}
\end{figure*}
\subsection{LLM-based Agent Simulation Evaluation}  \label{sec:agent}
\paragraph{\textit{Motivation}:} The LLM-based agent is the core component of Agent4Edu. Exploring whether the agent can truly simulate human learners' practice response is crucial for enhancing intelligent educational systems. We evaluate the effectiveness of the generative agent, including response simulation, exercises' knowledge understanding, zero-shot simulations, and ablation experiments.
\paragraph{Learner Simulation Evaluation}
The agent aims to generate simulated learner response data that closely approximates real responses. To validate the effectiveness of the simulation, we compare it with two traditional supervised simulation methods, including DAISIM~\cite{zhao2023simulating} and KES~\cite{liu2019exploiting}. Additionally, to enrich our baseline for a compelling comparison, we include several Knowledge Tracing (KT) models, such as DKVMN~\cite{zhang2017dynamic}, EERNN (with Markov)~\cite{su2018exercise} and SAKT~\cite{pandey2019self}, which are similar to the learner simulator in terms of response prediction.

In the experimental setup, each learner's records are divided into a 90\% training set and a 10\% test set. Each baseline model which is data-driven is trained on the training data, with the last 20\% records of each learner’s training data used for model validation. The agent has access to all training data to generate profiles and update its memory through reflection. During the testing phase, each trained baseline is tasked with predicting learners' binary responses (correct or incorrect) to unseen exercises in the test data. For our generative agent, exercises from the test data are sequentially sent to it, and it performs the designed three actions to solve them. If the agent rejects an exercise due to its difficulty, we label its response as an ``incorrect answer''.
The evaluation metrics are selected from two perspectives. Firstly, we use accuracy (ACC) and F1-score to measure prediction accuracy. Secondly, we assess the similarity between the simulated and real data distributions using ROUGE-3, inspired by~\cite{zhao2023simulating}.
We repeatedly run each baseline model five times in the same setups and the Table~\ref{tab:prediction} reports the average scores.

The experimental results indicate that Agent4Edu (GPT-3.5-turbo) demonstrates strong competitiveness compared to the supervised baselines, particularly in terms of ACC and F1-score. This suggests that the LLM-based agent has the potential to generate learner response data that closely resembles real-world datasets. Furthermore, among the baselines, EERNN performs exceptionally well by effectively modeling the exercise content as supplementary clues. Finally, an exploratory simulation conducted using Agent4Edu (GPT-3.5-turbo and GPT-4) on a subset of data with 100 learners shows that they enable the simulated distribution to closely approach the real distribution. Among them, GPT-4 performs better in terms of ACC and F1-score.

Additionally, we evaluate whether the simulated distribution of the agent’s practice success rate aligns with the actual distribution of learner data. We use the real response success rate as the ground truth and then replace the corresponding responses in the real sequence with the predicted responses from the test data to calculate the agent’s simulated success rate, as shown in Figure~\ref{fig:circle} (a). The comparison between the ground truth values and the agent’s results indicates that the simulated data effectively captures the learners’ practice patterns related to success rate.

\begin{table}[t]
    \centering
    \small
    \setlength{\tabcolsep}{2pt}
    \renewcommand{\arraystretch}{1}
    \begin{tabular}{l|ccc} \hline 
         Model &  ACC $\uparrow$ &  F1-score $\uparrow$ & ROUGE-3 $\uparrow$ \\ \hline 
         KES & 50.11 & 58.32 & 25.77 \\  
         DKVMN & 64.39 & 76.70 & 37.24 \\  
         EERNN & \underline{65.72} & {76.06} & \textbf{43.55} \\ 
         SAKT & {65.52} & \underline{78.33} & 31.09 \\ 
         DAISIM & 65.63 & 78.25 & 31.72 
         \\ \hline 
         Agent4Edu (GPT-3.5-turbo) & \textbf{66.70} & \textbf{79.84} & \underline{37.97} \\ \hline\hline  
         Agent4Edu (GPT-3.5-turbo)$_{100}$ & {65.40} & {78.72} & {35.14} \\ \hline 
             Agent4Edu (GPT-4)$_{100}$ & {66.51} & {79.53} & 34.86 \\ \hline
    \end{tabular}
    \caption{Prediction scores (\%) on evaluating simulation performance. The best results are bold, the second-best results are marked by an underline, and $\uparrow$ means the higher score the better performance, the same as below. Agent4Edu$_{100}$ indicates a basic exploratory on simulating 100 learners.}
    \label{tab:prediction}
\end{table}

\begin{table}[t]
    \centering
    \begin{tabular}{l|c}
        \toprule
         Model & ACC $\uparrow$ \\
        \midrule
        Agent4Edu (GPT-3.5-turbo) & 73.88 \\ \hline 
        Agent4Edu (GPT-3.5-turbo)$_{100}$ & 74.57 \\ \hline 
        Agent4Edu (GPT-4)$_{100}$ & 82.43 \\
        \bottomrule
    \end{tabular}
    \caption{The ACC of knowledge prediction.}
    \label{tab:question_understanding}
\end{table}

\paragraph{Understanding Exercise-related Knowledge}
To evaluate whether the agent understands a specific exercise, the agent is tasked with generating the knowledge concept tested by the exercise. Specifically, we create a candidate list containing one actual knowledge concept related to the exercise and two random knowledge concepts unrelated to the exercise. The agent must then select the relevant knowledge concept from this list based on its understanding (detailed prompts are provided in Appendix~C). We use ACC as the metric to evaluate the agent's knowledge predictions for all exercises in the test set, treating it as a binary classification task. This section uses the same agents from the section ``Learner Simulation Evaluation''. The experimental results presented in Table~\ref{tab:question_understanding}, indicate that all the agents can correctly identify the knowledge being tested in most practice exercises. This demonstrates the strong human-like ability and rich knowledge of LLMs to comprehend exercises.
Furthermore, under the same conditions with 100 learners, the agent with GPT-4 is more accurate than the one with GPT-3.5-turbo, indicating that GPT-4 has a stronger semantic understanding ability compared to GPT-3.5-turbo.

\paragraph{Zero-shot Simulation}
Zero-shot simulation presents a significant challenge in real-world applications, particularly when learners are in a cold-start situation where their response data is unavailable. This limitation restricts the applicability of previous simulation models. To validate the zero-shot simulation capability of the agent, we initialize 10 agents with randomly generated profiles and have them sequentially answer 10 randomly selected exercises with IDs $\{ 120, 250, 113, 1330, 568, 1881, 771, 593, 1,595\}$. In this zero-shot scenario, we disable the corrective reflection mechanism and tools, due to the absence of learner response data. The summary reflection remains usable for the agent.
Three GPT-3.5-turbo models, with a temperature parameter of 0.5, act as annotators, tasked with evaluating whether each simulated record (including exercise answers and practice summaries) is written by a real human. Records deemed to be human-written are labeled as ``Agent4Edu Win'', non-human records are labeled as ``Lose'', and ambiguous records are labeled as ``Tie''.

The results depicted in Figure~\ref{fig:circle}~(b) indicate that the agent’s performance in summarization is closely aligned with the real human responses, making differentiation between the two challenging. However, the agent exhibits certain limitations in answering exercise tasks compared to summarization tasks, primarily due to the complexity of reasoning required to solve exercises.

\paragraph{Ablation Study}
We conduct ablation studies to evaluate the impact of key components within the GPT-3.5-turbo-powered agent. The results illustrated in Figure~\ref{fig:circle}~(c) show the accuracy of the agent's exercise-related knowledge prediction and response prediction under various conditions: without the profile module (w/o prof), without the memory module (w/o mem), without the memory enhancement (w/o enh), without the memory forgetting (w/o fgt), and without reflection (w/o ref). These findings confirm the effectiveness of each component in improving the agent’s predictive performance on learners' response data. However, the ablation experiments indicate that the impact on knowledge prediction is not significant. This can be attributed to the fact that the original GPT-3.5-turbo model already possesses a substantial amount of knowledge, which is sufficient to support exercise comprehension.

\subsection{Personalized Learning}
\paragraph{\textit{Motivation}}
The primary objective of Agent4Edu is to comprehensively and accurately evaluate personalized learning algorithms and use the generated data to enhance their effectiveness. We aim to validate this objective from two perspectives: (1) through an agent-based multifaceted evaluation of personalized learning services, and (2) by assessing the potential improvements in personalized learning algorithms based on the simulated data.

\begin{table}[t]
    \centering
    \begin{tabular}{l|ccc}
        \toprule
         Model & satisfaction & \;\;AoD\;\; & \;\;gain\;\; \\
        \midrule
        FSI & \underline{39} & \textbf{70} & $\textbf{48}$ \\ 
        KLI & \underline{39} & ${66}$ & 43 \\
        MAAT & $\textbf{42}$ & \underline{68} & \underline{45} \\
        \bottomrule
    \end{tabular}
    \caption{Multifaceted evaluations of CAT strategies.}
    \label{tab:evaluation}
\end{table}

\paragraph{Multifaceted Evaluation}
Human learners have multifaceted evaluations of different personalized learning services, such as whether the recommended task difficulty is too challenging. Assuming that a generative agent can accurately simulate the behavior of real learners, its evaluation of personalized algorithms tends to align with human evaluations. We utilize the Computerized Adaptive Testing (CAT) which aims to estimate learners’ ability or knowledge proficiency with minor exercises, as the experimental environment, including FSI~\cite{lord2012applications}, KLI~\cite{chang1996global}, and MAAT~\cite{bi2020quality}. We use 100 randomly initialized agents to generate virtual data for pretraining the cognitive diagnosis model (i.e., the IRT model~\cite{baker2001basics}) in the CAT algorithm for learner evaluation. Based on this, each adaptive algorithm iterates through 10 rounds to recommend 100 randomly initialized agents (zero-shot simulation), with one exercise recommendation per round.
Upon the conclusion of personalized testing, each agent is required to evaluate each CAT algorithm. To achieve this, we design three evaluation metrics, including satisfaction, appropriateness of difficulty (AoD), and whether there was any gain.
Table~\ref{tab:evaluation} presents a comprehensive evaluation of various strategies, where the element in the $i$-th row and $j$-th column represents the number of agents that consider the corresponding CAT algorithm $i$ to meet the metric $j$. Clearly, the agent demonstrates higher satisfaction in recommending MAAT. This observation aligns with the common understanding in the research community that MAAT considers both the difficulty of exercises and the diversity of knowledge~\cite{bi2020quality}, making the overall service more reasonable. Additionally, FSI focuses on recommending exercises that are moderately difficult and likely to provide gain. These findings highlight the LLM-powered agent's fine-grained evaluation level for learning algorithms.

\paragraph{Personalized Learning Algorithm Improvement}
We investigate whether the simulated data generated by Agent4Edu can enhance personalized learning algorithms. We select CAT as our personalized learning assessment task due to their representativeness in intelligent education. If the generated data can improve the performance of CAT models, it will indicate the effectiveness of our proposed Agent4Edu.

To set up, we select 60\% of the learners' data from EduData to train the cognitive diagnosis models (i.e., the IRT model) for learner evaluation in CAT algorithms (i.e., FSI~\cite{lord2012applications}, KLI~\cite{chang1996global}, and MAAT~\cite{bi2020quality}). The remaining 40\% of learners' data is used to test the CAT models. Furthermore, for each learner in the test data, we simulate their responses to 20 randomly selected unseen exercises based on their profiles. Using this strategy, we generate simulated learner data, which are then merged with the training data from the original EduData to form the augmented dataset, EduData+. We train the IRT model in each CAT model using both the original EduData and EduData+, and then evaluate each CAT strategy by recommending 5 and 10 test exercises for each learner.

Table~\ref{tab:improvement} lists the IRT prediction performance after retraining on the testing records via CAT, where F1-score represents scores on EduData, and F1-score+ represents scores on EduData+. The results demonstrate that CAT strategies can be effectively enhanced with the assistance of Agent4Edu. This suggests that Agent4Edu is capable of generating high-quality learner response data, even with randomly initialized agents (in zero-shot scenarios), thereby enriching the provided dataset.

\begin{table}[t]
    \centering
    \setlength{\tabcolsep}{3pt} 
    \renewcommand{\arraystretch}{1} 
    \begin{tabular}{l|ccc} \hline 
    & \multicolumn{3}{c}{Testing length is 5} \\
    \hline
         Model & F1-score &  F1-score+ & Imp. \\ \hline 
         FSI & 80.11 & 82.39 & +2.28  \\  
         KLI & 79.45 & 81.84 & +2.39 \\
         MAAT & 81.77 & 81.97 & +0.20 \\ \hline\hline
         & \multicolumn{3}{c}{Testing length is 10} \\ \hline
         Model & F1-score  & F1-score+ & Imp.\\ \hline 
         FSI & 81.10 & 82.51 & +1.41  \\  
         KLI & 80.63 & 82.82 & +2.19 \\
         MAAT & 81.71 & 81.88 & +0.17 \\ \hline
    \end{tabular}
    \caption{The improvement of CAT services.}
    \label{tab:improvement}
\end{table}

\section{Conclusion}
In this paper, we introduce Agent4Edu, an innovative personalized learning simulator that leverages LLM-powered generative agents to simulate learners' response data, as well as detailed problem-solving behaviors. Our generative agents are equipped with learner Profile, Memory and Action modules specifically tailored for personalized learning scenarios. These agents exhibit human-like choosing, understanding, analyzing and answering exercises, which accurately predict their future responses.
Additionally, the generative agent can interact with personalized learning environments to evaluate and enhance intelligent services. Through comprehensive and meticulous evaluation, we explore the strengths and weaknesses of Agent4Edu, emphasizing the consistency and discrepancies in practice behaviors observed between agents and learners.
In the future, we plan to research multi-learner agent cooperation and multi-modal practice solutions using generative agents. We hope that our research will provide new insights into the field of intelligent education.

\section*{Acknowledgements}
This research was partially supported by grants from the National Natural Science Foundation of China (No.62337001, 62477044), the Key Technologies R \& D Program of Anhui Province (No. 202423k09020039), and the Fundamental Research Funds for the Central Universities.

\bibliography{aaai25}

\newpage

\appendix

\renewcommand{\thesection}{\Alph{section}}
\renewcommand{\thesubsection}{\thesection.\arabic{subsection}}

\appendixheading

\appendixsection{Basic Setups}
\label{app:basic_setup}

\subsection{Split of Profile Traits}
Using the traits of practice styles and cognitive factors defined in the subsection ``Learner Profile Module'', we categorize learners based on the strength of each feature. Specifically, we classify each learner's \textbf{learning activity} \( P_{act}^{u} \) and \textbf{diversity} \( P_{div}^{u} \) as either ``high'' or ``low'', depending on whether the trait's value is above or below the median across all learners. Additionally, we divide each learner's \textbf{success rate} \( P_{suc}^{u} \) and \textbf{ability} \( P_{ab}^{u} \) into ``low'', ``medium'', or ``high'' categories, in equal proportions, based on the ascending order of all learners.


\subsection{Split of Knowledge Proficiency}
Knowledge proficiency is a dynamic cognitive factor generated by the DNeuralCDM tool, which tracks a learner's knowledge proficiency over time by analyzing each response record. We categorize each learner's \textbf{knowledge proficiency} at each step into ``high'', ``medium'' or ``low'' categories, in equal proportions, based on ascending order of all learners.

\subsection{The Workflow of Initializing Agents based on Learner Response Data}
\begin{algorithm}[ht]
\caption{The Workflow of Initializing LLM-based Agents by Simulating Response Data}
\label{alg:llm-agent-simulation}
\begin{algorithmic}[1]
\REQUIRE Response data for each learner $u\in U$
\STATE Initialize \textit{profile}, \textit{factual memory}, \textit{short-term memory}, and \textit{long-term memory} as empty.
\FOR{each learner $u \in U$}
    \STATE Generate \textit{learner profile} based on learner $u$'s data $l_u$.
    \FOR{each record $(e_i, c_{e_i}, y_{u,e_i}) \in l_u$}
        \STATE Add this record to \textit{factual memory} and update each counter $f_{u,i}$.
        \STATE Update \textit{short-term memory} using recent records.
        \STATE Retrieve reinforced facts in factual memory as important facts.
        \STATE Generate \textit{knowledge proficiency} using the DNeuralCDM tool.
        \STATE Summarize reflections to generate learning status.
        \STATE Generate a prompt for the agent to perform actions.
        \IF{the agent's response to $e_{i+1}$ requires corrective reflection}
            \STATE Update \textit{learning status} and generate a new prompt until the action is correct.
        \ENDIF
    \ENDFOR
\ENDFOR
\end{algorithmic}
\end{algorithm}

\appendixsection{Tools} \label{app:tool}
\subsection{IRT Model}
\paragraph{\textbf{Basic Description}}
IRT~\cite{baker2001basics} originated in educational psychology and is used to model learners' abilities by analyzing their response data. Generally, IRT assumes that a learner's ability is stable and positively correlated with their responses to exercises. Therefore, by fitting the learners' response data, the ability parameters can be optimized. 
Given the response data $\{l_u\}_{u\in U}$ of each learner $u \in U$, the optimization objective of IRT is as follows:
\begin{equation}
\begin{split}
    \mathcal{L}_{IRT} &= \sum_{u\in U} \sum_{i\in |l_u|} \ell(y_{u,e_i}, \hat{y}_{u,e_i}), \\
    \hat{y}_{u,e_i} &= \frac{1}{1 + \exp(-( \theta_u - b_{e_i} ))},
\end{split}
\end{equation}
where $l(\cdot)$ is usually the binary cross-entropy to minimize the gap between the predicted probability $\hat{y}_{u,e_i}\in[0,1]$ that learner $u$ correctly answers $e_{i}$ and the ground-truth response $y_{u,e_i} \in \{0,1\}$. $\theta_u \in [0,1]$ is learner $u$'s ability parameter and $b_{e_i} \in [0,1]$ denotes exercise $e_j$'s difficulty parameter. By joint training, $\theta_u$ is optimized to capture the learner's problem-solving ability.

\paragraph{\textbf{Training}}
The IRT model is trained using the training data (as detailed in the ``Learner Simulation Evaluation'' section in 4.1), and the version achieving the highest ACC score on the validation set is selected as the tool.

\paragraph{\textbf{Inference}}
Using the optimized IRT model, the agent $agent_u$ calls it as an external tool to obtain the ability parameter $\theta_u \in [0,1]$ for learner $u$. This ability parameter is then further divided into specific levels to complete the learner profile, as detailed in Appendix~\ref{app:basic_setup}.

\subsection{DNeuralCDM}

\paragraph{\textbf{Basic Description}}
DNeuralCDM~\cite{wang2023dynamic} dynamically traces the evolution of learners' knowledge proficiency throughout their practice sessions. Unlike the IRT model, which analyzes learner exercise data in a non-sequential manner, DNeuralCDM processes each learner $u$'s response record $l_u$ sequentially, adhering to the chronological order of practice. Formally, given the response record $(e_t, c_{e_t}, y_{u,e_t})$ at time \( t \) (where \( t = 1, \dots, |l_u|-1 \)), DNeuralCDM first calculates the learner's knowledge proficiency on each knowledge concept at step $t$, and then predicts the probability of correctly answering all questions at the next time step. The optimization objective of DNeuralCDM, akin to that of IRT, is to fit the learner's performance $y_{u,e_{t+1}}$ at time \( (t+1) \). The DNeuralCDM model is computed as:
\begin{equation}
\begin{split}
    \mathcal{L}_{\text{DNeuralCDM}} &= \sum_{u \in U} \sum_{t \in [1, |l_u|-1]} \ell(y_{u, e_{t+1}}, \hat{y}_{u, e_{t+1}}), \\
    \hat{y}_{u, e_{t+1}} &= f(h_{u,t}, y_{u, e_t}),
\end{split}
\end{equation}
where \( l(\cdot) \) represents the binary cross-entropy loss function used in the DNeuralCDM model. This function minimizes the discrepancy between the actual performance of learner \( u \) on exercise \( e_{t+1} \) (denoted by \( y_{u, e_{t+1}} \in \{0,1\}\)) and the model's predicted performance \( \hat{y}_{u, e_{t+1}} \in [0,1] \). The function \( f(h_{u,t}, y_{u, e_t}) \) computes the predicted performance at time $(t+1)$ based on the learner's knowledge state \( h_{u,t} \in \mathbb{R}^{|C|} \) at time \( t \) and the actual performance \( y_{u, e_t} \) at the same time. \( C \) denotes the set of knowledge concepts covered by all exercises \( E \). The detailed structure of \( f(\cdot) \) is not provided here. For a comprehensive explanation, please refer to the original paper.

\paragraph{\textbf{Training}}
The DNeuralCDM model is trained using the training data (as detailed in the ``Learner Simulation Evaluation'' section in 4.1), and the version achieving the highest ACC score on the validation set is selected as the tool.

\paragraph{\textbf{Inference}}
Given the optimized DNeuralCDM model, it is utilized by the agent \( {agent}_u \) to compute the knowledge proficiency parameter \( h_{u,t} \in \mathbb{R}^{|C|} \) for learner \( u \) at step \( t \). The \( c \)-th element of \( h_{u,t} \) denotes learner \( u \)'s proficiency in knowledge concept \( c \). At each step \( t \), the agent retrieves only the proficiency level associated with the knowledge concept tested by exercise \( e_t \). This proficiency parameter is subsequently partitioned into specific levels to complete the learner profile, as detailed in Appendix~\ref{app:basic_setup}. 

\appendixsection{Prompts} \label{app:prompt}
\subsection{Prompt Designing} \label{app:prompt_design}

In this section, we use {placeholder} to denote a placeholder that need to be filled. The possible
values of these placeholders can be found in Section~\ref{app:placeholder}.

\noindent$\bullet$ The prompt for creating a learner's profile in the \textbf{Learner Profile Module} is as follows:

\begin{tcolorbox}[
    enhanced,
    colback=lightpurple,
    colframe=mypurple,
    coltitle=white,
    fonttitle=\bfseries,
    title=Learner Profile Generation,
    breakable
]
You are a high school student engaging in self-directed exercising on an online learning platform. During online study, you exhibit \{\textit{activity}\} activity, which means \{\textit{activity\_description}\}. You have \{\textit{diversity}\} knowledge diversity, which means \{\textit{diversity\_description}\}. The knowledge concept you practice most often is: \{\textit{preference}\}. Your success rate is \{\textit{success\_rate}\}. You possess \{\textit{ability}\} analytical and problem-solving skills.

The information above is your \# profile \#.
\end{tcolorbox}

\noindent$\bullet$ The knowledge relation judgment in the \textbf{Factual Memory Module} is implemented by the RCD-based KCG tool rather than an LLM prompt. Given two response records, Agent4Edu compares their knowledge concepts using the constructed knowledge concept graph. If the concepts are related in the graph, the corresponding factual memory is reinforced.

\noindent$\bullet$ The prompt for reflection in the \textbf{Memory Module} is designed as follows:

\begin{tcolorbox}[
    enhanced,
    colback=lightpurple,
    colframe=mypurple,
    coltitle=white,
    fonttitle=\bfseries,
    title=Memory Reflection,
    breakable
]
\textit{\# If corrective feedback exists, it is placed before the following instruction.}

You should directly output your reflection and summarize your \# Learning Status \# within 500 words based on your \# profile \#, \# short-term memory \#, \# long-term memory \# and previous \# Learning Status \#. Do not output any other information.
\end{tcolorbox}

\noindent$\bullet$ The prompt for taking actions for the agent, corresponding to the \textbf{Action Module}, is designed as follows:
\begin{tcolorbox}[
    enhanced,
    colback=lightpurple,
    colframe=mypurple,
    coltitle=white,
    fonttitle=\bfseries,
    title=Taking Actions (A complete Workflow),
    breakable
]
\textbf{Learner profile:}

You are a high school student engaging in self-directed exercising on an online learning platform. During online study, you exhibit \{\textit{activity}\} activity, which means \{\textit{activity\_description}\}. You have \{\textit{diversity}\} knowledge diversity, which means \{\textit{diversity\_description}\}. The knowledge concept you practice most often is: \{\textit{preference}\}. Your success rate is \{\textit{success\_rate}\}. You possess \{\textit{ability}\} analytical and problem-solving skills.

The information above is your \# profile \#.

\textbf{Short-term Memory:}

I will give you recent practice records as \# Recent Facts \# below.

Record1: You \{\textit{result}\} answered an exercise.

\noindent{}- \# Textual Content \#: \{\textit{question\_content}\}

\noindent{}- \# Knowledge Concept \#: \{\textit{knowledge\_concept}\}

$\cdots$

The information above is your \# short-term memory \#.

\textbf{Long-term Memory:}

I will give you important reinforced records as \# Reinforced Facts \# below.

Record1: You \{\textit{result}\} answered an exercise.

\noindent{}- \# Textual Content \#: \{\textit{question\_content}\}

\noindent{}- \# Knowledge Concept \#: \{\textit{knowledge\_concept}\}

$\cdots$

Your current \# Knowledge Proficiency \# is:

\noindent{}- \{\textit{concept}\}: \{\textit{proficiency\_level}\} (\{\textit{proficiency\_value}\})

$\cdots$

Your current \# Learning Status \# is summarized as:

\{\textit{status}\}

The information above is your \# long-term memory \#.

\textbf{Actions:}

Currently, you start to answer the recommended exercise. Its content information is as follows:

\# Textual Content \#: \{\textit{question\_content}\}

\# Options \#: \{\textit{question\_options}\}

\# Reference Answer \#: \{\textit{reference\_answer}\}

\# Analysis \#: \{\textit{question\_analysis}\}

To answer this exercise, please complete the following four tasks in sequence:

\begin{itemize}
\item Task 1 is to decide whether to attempt the recommended problem based on your ability in Profile and knowledge proficiency in Long-term Memory. If you consider the problem too difficult, output ``No''; otherwise output ``Yes''. Regardless of your choice, the subsequent tasks will still be executed.

\item Task 2 is to choose one knowledge concept tested by this exercise from the following three options:

\noindent{}- \{\textit{concept\_1}\}

\noindent{}- \{\textit{concept\_2}\}

\noindent{}- \{\textit{concept\_3}\}

Only output the knowledge concept and do not output any other information for Task 2.

\item Task 3 is to design a short problem-solving idea for this question based on your profile, memory and learning status, and then give a final answer. Your response should align with your profile, memory, and past performance.

\item Task 4 is to estimate whether you can correctly solve this problem based on your profile, learning records, learning status, and problem-solving idea. If you can correctly solve it, answer ``Yes''; otherwise answer ``No''.
\end{itemize}

Output exactly in this format:

Task1: $<$answer for task1$>$

Task2: $<$answer for task2$>$

Task3: $<$answer for task3$>$

Task4: $<$answer for task4$>$

\end{tcolorbox}

\noindent$\bullet$ The prompt for \textbf{Corrective Reflection} to address incorrect actions is designed as follows:

\begin{tcolorbox}[
    enhanced,
    colback=lightpurple,
    colframe=mypurple,
    coltitle=white,
    fonttitle=\bfseries,
    title=Corrective Reflection,
    breakable
]

\textit{\# If the agent outputs a wrong knowledge concept in Task 2, the following corrective feedback is added:}

The knowledge tested by this question is \{\textit{test\_knowledge}\} but you wrongly think the knowledge is \{\textit{pre\_knowledge}\}.

\textit{\# If the agent predicts that it cannot solve a problem correctly, but the real performance is correct, the following corrective feedback is added:}

You thought you could not solve this problem correctly, but in fact, you will solve it correctly.

\textit{\# If the agent predicts that it can solve a problem correctly, but the real performance is wrong, the following corrective feedback is added:}

You thought you could solve this problem correctly, but in fact, you do not answer it correctly.

You should directly output your reflection and summarize your \# Learning Status \# within 500 words based on your \# profile \#, \# short-term memory \#, \# long-term memory \# and previous \# Learning Status \#. Do not output any other information.

\end{tcolorbox}

\subsection{Placeholders in Prompts} \label{app:placeholder}
In this section, we present examples of placeholders in prompts.
\begin{tcolorbox}[
    enhanced,
    colback=lightlakeblue,
    colframe=mylakeblue,
    coltitle=white,
    fonttitle=\bfseries,
    title=Placeholder Examples,
    breakable
]
\textbf{\{\textit{activity}\}: \{\textit{activity\_description}\}}
\begin{itemize}
\item high: you maintain a high level of online exercise activity and practice frequently.

\item low: you practice less regularly and with lower enthusiasm.
\end{itemize}

\textbf{}

\textbf{\{\textit{diversity}\}: \{\textit{diversity\_description}\}}
\begin{itemize}
\item high: you explore diverse knowledge categories.

\item low: you focus on limited knowledge categories.
\end{itemize}

\textbf{}

\textbf{\{\textit{preference}\}}
\begin{itemize}
\item The knowledge concept practiced most often by the learner.
\end{itemize}

\textbf{}

\textbf{\{\textit{success\_rate}\}}
\begin{itemize}
\item high.
\item medium.
\item low.
\end{itemize}

\textbf{}

\textbf{\{\textit{ability}\}} (obtained by IRT model tool \includegraphics[height=0.5cm]{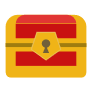})
\begin{itemize}
\item good.

\item common.

\item poor.
\end{itemize}

\textbf{}

\textbf{\{\textit{result}\}}
\begin{itemize}
\item rightly.
\item wrongly.
\end{itemize}

\textbf{}

\textbf{\{\textit{knowledge\_concept}\}}
\begin{itemize}
\item The knowledge concept tested by the exercise.
\end{itemize}

\textbf{}

\textbf{\{\textit{proficiency\_level}\}} (obtained by DNeuralCDM tool \includegraphics[height=0.5cm]{tool.png})
\begin{itemize}
\item good.

\item common.

\item poor.

\item unknown.
\end{itemize}

\textbf{}

\textbf{\{\textit{proficiency\_value}\}}
\begin{itemize}
\item The numerical proficiency value produced by DNeuralCDM. If unavailable, the value is shown as unknown.
\end{itemize}

\textbf{}

\textbf{\{\textit{status}\}}
\begin{itemize}
\item The latest learning status summarized by memory reflection.
\end{itemize}

\textbf{}

\textbf{\textit{\{test\_knowledge\}}}
\begin{itemize}
\item The true knowledge concept tested by the current practice.
\end{itemize}

\textbf{}

\textbf{\textit{\{pre\_knowledge\}}}
\begin{itemize}
\item The knowledge concept predicted by the generative agent for the current practice.
\end{itemize}

\textbf{}

\textbf{\textit{\{question\_content\}}}
\begin{itemize}
\item The textual content of the current practice.
\end{itemize}

\textbf{}

\textbf{\textit{\{question\_options\}}}
\begin{itemize}
\item The options of the current practice, if available.
\end{itemize}

\textbf{}

\textbf{\textit{\{reference\_answer\}}}
\begin{itemize}
\item The reference answer of the current practice, if available.
\end{itemize}

\textbf{}

\textbf{\textit{\{question\_analysis\}}}
\begin{itemize}
\item The reference analysis of the current practice, if available.
\end{itemize}

\textbf{}

\textbf{\textit{\{concept\_1\}}, \textit{\{concept\_2\}}, \textit{\{concept\_3\}}}
\begin{itemize}
\item Three candidate knowledge concepts, including the true concept and two sampled distractors.
\end{itemize}

\end{tcolorbox}

\appendixsection{A Case Study of Agent4Edu for CAT} \label{app:example_Agent4Edu}
In this section, we present an example demonstrating the interaction between the agent $agent_u$ and the CAT algorithm. The example follows the same prompt format used in the implementation. The agent receives its profile, short-term memory, long-term memory, and the recommended exercise, then completes the four action tasks and updates its learning status through reflection when necessary.

\begin{tcolorbox}[
    enhanced,
    colback=lightcat,
    colframe=mycat,
    coltitle=white,
    fonttitle=\bfseries,
    title=The CAT Algorithm,
    breakable
]
\vspace{0.1cm}
\begin{minipage}{1cm}
    \includegraphics[height=1cm]{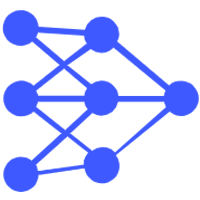}
\end{minipage}%
\hspace{0.1cm} 
\begin{minipage}{\dimexpr\linewidth-1.2cm\relax}
    The CAT Algorithm recommends one exercise for the Learner Agent $agent_u$ at each step of practice based on its current response data.
\end{minipage}

\vspace{0.2cm}

\noindent\rule{\linewidth}{0.3pt}

\vspace{0.2cm}

\begin{minipage}{1cm}
    \includegraphics[height=1cm]{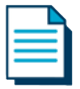}
\end{minipage}%
\hspace{0.1cm} 
\begin{minipage}{\dimexpr\linewidth-1.2cm\relax}
    The basic description of the recommended exercise is as below:
\end{minipage}

\vspace{0.2cm}

\textcolor{blue}{[Key]:} Textual Content

\textcolor{blue}{[Value]:}
Will the graph of the function $y=a^x-3$ always pass through the point $(1, a)$?
\end{tcolorbox}

\begin{tcolorbox}[
    enhanced,
    colback=lightprompt,
    colframe=myprompt,
    coltitle=white,
    fonttitle=\bfseries,
    title=Prompts for the Generative Agent $agent_u$,
    breakable
]
\textbf{Learner Profile:}

You are a high school student engaging in self-directed exercising on an online learning platform. During online study, you exhibit \textcolor{myprompt}{\textit{low}} activity, which means \textcolor{myprompt}{\textit{you practice less regularly and with lower enthusiasm}}. You have \textcolor{myprompt}{\textit{low}} knowledge diversity, which means \textcolor{myprompt}{\textit{you focus on limited knowledge categories}}. The knowledge concept you practice most often is: \textcolor{myprompt}{\textit{Geometry}}. Your success rate is \textcolor{myprompt}{\textit{low}}. You possess \textcolor{myprompt}{\textit{common}} analytical and problem-solving skills.

The information above is your \# profile \#.

\textbf{Short-term Memory:}

I will give you recent practice records as \# Recent Facts \# below.

Record1: You rightly answered an exercise.

\noindent{}- \# Textual Content \#: In a right triangle, one of the legs is 6 units long, and the other leg is 8 units long. What is the length of the hypotenuse?

\noindent{}- \# Knowledge Concept \#: Geometry

Record2: You wrongly answered an exercise.

\noindent{}- \# Textual Content \#: Given the function \( f(x) = 2x^2 - 3x + 5 \), what is the value of \( f(2) \)?

\noindent{}- \# Knowledge Concept \#: Function

Record3: You wrongly answered an exercise.

\noindent{}- \# Textual Content \#: If \( f(x) = 3x + 4 \) and \( g(x) = x^2 - 2x + 1 \), find the value of \( (f \circ g)(2) \), where \( (f \circ g)(x) \) represents the composition of the functions \( f \) and \( g \).

\noindent{}- \# Knowledge Concept \#: Function

Record4: You rightly answered an exercise.

\noindent{}- \# Textual Content \#: Consider the function \( h(x) = \frac{5x - 1}{x + 2} \). Find the value of \( h(-1) \).

\noindent{}- \# Knowledge Concept \#: Function

The information above is your \# short-term memory \#.

\textbf{Long-term Memory:}

I will give you important reinforced records as \# Reinforced Facts \# below.

Record1: You wrongly answered an exercise.

\noindent{}- \# Textual Content \#: Given the function \( f(x) = 2x^2 - 3x + 5 \), what is the value of \( f(2) \)?

\noindent{}- \# Knowledge Concept \#: Function

Your current \# Knowledge Proficiency \# is:

\noindent{}- Geometry: good (0.7800)

\noindent{}- Function: poor (0.2100)

Your current \# Learning Status \# is summarized as:

You have practiced both geometry and function-related exercises. Your recent records indicate that you can solve some function questions, but you still make mistakes on several function-related problems. You should pay more attention to function expressions and substitution steps.

The information above is your \# long-term memory \#.

\textbf{Actions:}

Currently, you start to answer the recommended exercise. Its content information is as follows:

\# Textual Content \#: Will the graph of the function $y=a^x-3$ always pass through the point $(1, a)$?

\# Options \#:

\# Reference Answer \#: No.

\# Analysis \#: Substitute $x=1$ and $y=a$ into $y=a^x-3$. This gives $a=a-3$, which is impossible. Therefore, the graph does not always pass through $(1,a)$.

To answer this exercise, please complete the following four tasks in sequence:

\begin{itemize}
\item Task 1 is to decide whether to attempt the recommended problem based on your ability in Profile and knowledge proficiency in Long-term Memory. If you consider the problem too difficult, output ``No''; otherwise output ``Yes''. Regardless of your choice, the subsequent tasks will still be executed.

\item Task 2 is to choose one knowledge concept tested by this exercise from the following three options:

\noindent{}- Algorithm

\noindent{}- Function

\noindent{}- Force

Only output the knowledge concept and do not output any other information for Task 2.

\item Task 3 is to design a short problem-solving idea for this question based on your profile, memory and learning status, and then give a final answer. Your response should align with your profile, memory, and past performance.

\item Task 4 is to estimate whether you can correctly solve this problem based on your profile, learning records, learning status, and problem-solving idea. If you can correctly solve it, answer ``Yes''; otherwise answer ``No''.
\end{itemize}

Output exactly in this format:

Task1: $<$answer for task1$>$

Task2: $<$answer for task2$>$

Task3: $<$answer for task3$>$

Task4: $<$answer for task4$>$
\end{tcolorbox}

\begin{tcolorbox}[
    enhanced,
    colback=lightgptgreen,
    colframe=mygptgreen,
    coltitle=white,
    fonttitle=\bfseries,
    title=The Agent Answers,
    breakable
]
\includegraphics[height=1cm]{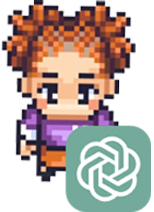}

\vspace{0.1cm}

Task1: Yes

Task2: Algorithm

Task3: Since the exercise contains the expression $y=a^x-3$, I would substitute the point $(1,a)$ into the function. When $x=1$, the function gives $y=a-3$, which does not equal $a$. Therefore, the graph does not always pass through $(1,a)$. The final answer is No.

Task4: Yes
\end{tcolorbox}

\begin{tcolorbox}[
    enhanced,
    colback=lightprompt,
    colframe=myprompt,
    coltitle=white,
    fonttitle=\bfseries,
    title=Prompts for the Generative Agent $agent_u$,
    breakable
]
\textbf{Corrective Reflection:}

The knowledge tested by this question is Function but you wrongly think the knowledge is Algorithm.

You should directly output your reflection and summarize your \# Learning Status \# within 500 words based on your \# profile \#, \# short-term memory \#, \# long-term memory \# and previous \# Learning Status \#. Do not output any other information.
\end{tcolorbox}

\begin{tcolorbox}[
    enhanced,
    colback=lightgptgreen,
    colframe=mygptgreen,
    coltitle=white,
    fonttitle=\bfseries,
    title=The Agent Answers,
    breakable
]
\includegraphics[height=1cm]{agent.png}

\vspace{0.1cm}

I incorrectly associated the exercise with Algorithm, but the main object in the problem is the function $y=a^x-3$ and whether its graph passes through a given point. This shows that I should focus more carefully on the tested mathematical concept before solving the problem. My current learning status is that I can sometimes reason through function problems, but I still confuse the tested concept when a problem contains unfamiliar expressions. I should strengthen my understanding of function graphs and substitution-based verification.
\end{tcolorbox}

\begin{tcolorbox}[
    enhanced,
    colback=lightprompt,
    colframe=myprompt,
    coltitle=white,
    fonttitle=\bfseries,
    title=Prompts for the Generative Agent $agent_u$,
    breakable
]
\textbf{Learner Profile:}

You are a high school student engaging in self-directed exercising on an online learning platform... (\textit{The detailed content is omitted.})

The information above is your \# profile \#.

\textbf{Short-term Memory:}

I will give you recent practice records as \# Recent Facts \# below... (\textit{The detailed content is omitted.})

The information above is your \# short-term memory \#.

\textbf{Long-term Memory:}

I will give you important reinforced records as \# Reinforced Facts \# below... (\textit{The detailed content is omitted.})

Your current \# Knowledge Proficiency \# is... (\textit{The detailed content is omitted.})

Your current \# Learning Status \# is summarized as:

I incorrectly associated the exercise with Algorithm, but the main object in the problem is the function $y=a^x-3$ and whether its graph passes through a given point. This shows that I should focus more carefully on the tested mathematical concept before solving the problem. My current learning status is that I can sometimes reason through function problems, but I still confuse the tested concept when a problem contains unfamiliar expressions. I should strengthen my understanding of function graphs and substitution-based verification.

The information above is your \# long-term memory \#.

\textbf{Actions:}

Currently, you start to answer the recommended exercise. Its content information is as follows:

\# Textual Content \#: Will the graph of the function $y=a^x-3$ always pass through the point $(1, a)$?

\# Options \#:

\# Reference Answer \#: No.

\# Analysis \#: Substitute $x=1$ and $y=a$ into $y=a^x-3$. This gives $a=a-3$, which is impossible. Therefore, the graph does not always pass through $(1,a)$.

To answer this exercise, please complete the following four tasks in sequence... (\textit{The detailed content is omitted.})

Output exactly in this format:

Task1: $<$answer for task1$>$

Task2: $<$answer for task2$>$

Task3: $<$answer for task3$>$

Task4: $<$answer for task4$>$
\end{tcolorbox}

\begin{tcolorbox}[
    enhanced,
    colback=lightgptgreen,
    colframe=mygptgreen,
    coltitle=white,
    fonttitle=\bfseries,
    title=The Agent Answers,
    breakable
]
\includegraphics[height=1cm]{agent.png}

\vspace{0.1cm}

Task1: Yes

Task2: Function

Task3: To determine whether the graph of $y=a^x-3$ always passes through $(1,a)$, substitute $x=1$ and $y=a$ into the function. This gives $a=a^1-3=a-3$, which implies $0=-3$. This is impossible, so the graph does not always pass through $(1,a)$. The final answer is No.

Task4: Yes
\end{tcolorbox}

\begin{tcolorbox}[
    enhanced,
    colback=lightcat,
    colframe=mycat,
    coltitle=white,
    fonttitle=\bfseries,
    title=The CAT Algorithm,
    breakable
]
\vspace{0.1cm}

\begin{minipage}{1cm}
    \includegraphics[height=1cm]{al.png}
\end{minipage}%
\hspace{0.1cm} 
\begin{minipage}{\dimexpr\linewidth-1.2cm\relax}
    The CAT algorithm selects a single exercise for the Learner Agent \( agent_u \) at each practice step, based on the most recent response data, which is continuously updated with the latest simulated responses from \( agent_u \).
\end{minipage}

\vspace{0.2cm}

\noindent\rule{\linewidth}{0.3pt}

\vspace{0.2cm}

\begin{minipage}{1cm}
    \includegraphics[height=1cm]{exercise.png}
\end{minipage}%
\hspace{0.1cm} 
\begin{minipage}{\dimexpr\linewidth-1.2cm\relax}
    The basic description of the recommended exercise is as below:
\end{minipage}

\vspace{0.2cm}

\textcolor{blue}{[Key]:} Textual Content

\textcolor{blue}{[Value]:} ...
\end{tcolorbox}

\end{document}